# Analysis of Integrating Blockchain Technologies into Multi-Agent Systems

Chelsea R. Woodward
Department of Science and Technology
Bournemouth University
Dorset, United Kingdom
s5432781@bournemouth.ac.uk

*Multi-Agent Systems, a division of Intelligent Systems diversely applied in multiple disciplines. Desired for their efficiency in solving complex problems at a low cost. However, identified vulnerabilities include system security, integrity, and identity management. Blockchain Technologies was chosen for analysis in providing a suitable solution due to features of transparency, encryption, and trust.*

*Keywords—Multi-Agent Systems, Blockchain Technologies, Intelligent Systems, Distributed Computing, Security, Trust*

## I. INTRODUCTION

Multi-Agent Systems (MAS) are a branch within the field of Distributed Artificial Intelligence (DAI), defined as a group of connected autonomous entities which interact in one environment to serve a particular purpose [1]. These intelligent systems are popularly used in domains such as Finance, Energy, and Electronic Healthcare (eHealth), desirable for their flexibility, affordability, and efficiency in solving complex problems [1, 2, 3, 4]. Conversely, system security, transparency, and co-ordination are characteristics of MAS identified as vulnerabilities, compromising system integrity [1]. Blockchain Technologies (BCT), a compatible choice for integration, have been proposed as a solution in recent research [3, 4, 5], however are lacking in completeness and actioning of this approach.

BCT is an area within Distributed Computing and can be described as distributed ledger-based systems which operate on a Peer-to-Peer basis. Most commonly associated with the successful implementation of cryptocurrency *Bitcoin,* but BCT is applicable in numerous domains, most of which overlap with those of MAS [6]. BCT is known best for its hash encryption feature, a method for providing high-level security to a system. Moreover, BCT is known to be capable of reducing risk, due to its decentralized nature, maintain co-ordination control, due to its consensus protocols, and provide trust due to its transparency in recording of all (trans)actions on the blockchain into an immutable ledger. Therefor incorporation of BCT into MAS would not only optimize security but has the potential to achieve an overall improved system with better performance.

This paper reviews the integration of BCT to the MAS framework as a method for resolving vulnerabilities present in MAS, considering the feasibility, advantages, and disadvantages. The structure of the paper is as follows: Section I introduced the topic area outlining the problem and proposed solution, Section II details MAS features, structure and technical challenges, Section III identifies the suitability of applying BCTs to MAS, Section IV entails an in-depth discussion on this integration, all of which is summarised with suggestions for future work in Section V.

## II. MAS STRUCTURE, FEATURES AND CHALLENGES

Agents have the ability to not only sense parameters within their environment, but can make decisions to perform actions, achieving goals which can alter the environment. Where each agent has a predetermined *knowledge*, capable of adding new information, and a set of *behaviours*, actions the agents are capable of performing. Agents of a *Multi-Agent System* can exchange knowledge and collaborate in performing tasks, thus allowing for solving complex problems. MAS have features of being autonomous, reactive, rational, and adaptive. Thereby **decisions are made based on the goal** of the MAS, alongside taking into consideration the agent knowledge, user needs, environmental factors and changes, history of completed actions as well as rationality and logic.

The management of MAS can be either *leaderless* or *leader-follow*. Wherein leader-follow one agent assumes the position of 'lead agent' and designates decisions and actions to the other agents. Nevertheless, MAS Systems are still considered to be decentralized [1], however, although this may avoid containing a 'Single Point of Failure', it does not protect against **faulty agents**. Faulty agents are ones who have become inadequate perhaps due to faulty sensors, actuators, or programming, which may in turn corrupt the other agents that are in collaboration with the faulty agent. This raises concerns for a trustworthy system. **Verification of agent identity and establishing trust** between agents is a core requirement and equally a vulnerability.

Co-ordination is a significant aspect of MAS, subdividing into issues including formation (tracking) [7, 8], achieving consensus [1, 4, 8], and agent synchronisation [1, 9]. Further challenges of MAS include agent reputation and identity management, including task allocation, organisation, and previously mentioned, security [1, 10, 11]. A depiction of defined weaknesses is condensed below in Figure 1 where all vulnerabilities affect data, and system, integrity.

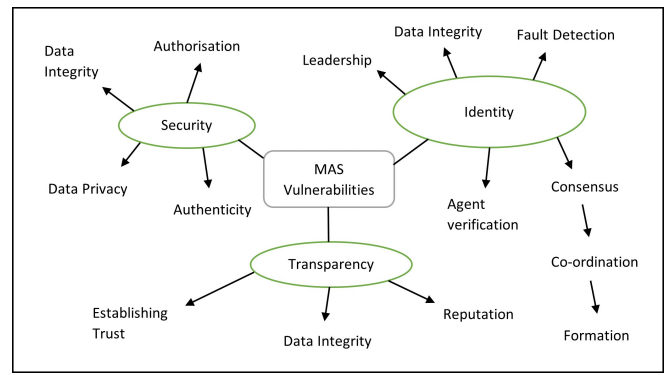

Figure 1 - A condenses depiction of defined MAS vulnerabilities

## III. BLOCKCHAIN TECHNOLOGIES, FEATURES AND SUITABILITY

Blockchains consist of a chain of blocks holding stored information, extending continuously when new blocks are added to the chain. BCTs act in a decentralised nature through an immutable distributed ledger. Therefor a history of all transactions is accessible, but not amendable. With each transaction made verification of the sender's identity along with independent validation is required, utilizing digital signatures, after which a hash encryption is generated [6]. The use of verification of digital signatures, 265-bit integers, in conjunction with the above mentioned **ensure trust, accountability and security in BCT-incorporated systems.**

Initialising BCTs involve the use of a smart contract, also referred to as the *Chaincode*. Smart Contracts consist of four main states, Creation, Deployment, Execution and Completion. Where contracts are *deployed* after the predetermined conditions are negotiated and defined during the *creation* stage. Transactions are then automatically *executed* once these criteria are met, upon *completion* of which all parties receive the new output(s). For example, within the energy domain, states and measurements of smart meters are recorded into and communicated through a blockchain. Where a smart contract is automatically executed to act between agents, once an energy block passes the consensus protocols and is successfully added to the chain, a function triggers a transaction for the appropriate funds to be released to the producer from the consumer (dependant on the energy value) [2]. One of the key benefits to a smart contract is that it is Immutable, were initial negotiations of contract terms and conditions cannot be later altered once agreed upon and the code has been written. This provides a transaction guarantee and authenticity of all involved individuals.

As previously mentioned, enhanced security can be achieved through BCT as it is a decentralised system, meaning no individual user may make alterations for the entire blockchain without approval. Additionally, BCT offers mechanisms such as encrypting transactions and requiring digital signatures, methods previously applied in protecting smart grids [2]. Moreover, permissioned blockchains act through a membership service, meaning all users, i.e., agents have each been identified and certificated which also plays a key role in introducing trust. Additionally, trust is directly achieved through the transparent nature of BCT, given all transactions and associated verifications on the blockchain are recorded into a history. In addition, BCT Consensus Protocols offer security mechanisms such as Proof of Work (PoW), a method to determine the verification of a node before any additions to the blockchain are made. Where decentralisation protects against possible *single points of failures,* and consensus protocols protect against attacks such as 'the 51% attack' [11], the mentioned security enhancements also offer system integrity and data protections, all important aspects of any system.

Consensus protocols are essential in achieving agreement within a distributed network. PoW consensus protocol is method of providing verification by means of a proof-based algorithm [6]. To summarize, consensus is reached regarding the addition of a new block when all miners achieve the same nonce value obtained after successful completion of a cryptographic puzzle, generated according to the target difficulty [6, 12, 13]. Where a nonce signifies *number used once*, and a miner is a participating node [6], and nodes can be considered as both host and server within the network that exchange information with one another [13]. Similarly, Proof of Stake (PoS) is a protocol which entails solving the same cryptographic puzzle, (SHA256 puzzle). However, unlike PoW protocol which requires large computational power, where typically the miner with the most computational power will generate the solution first, PoS focus is on the miner that is the highest stakeholder. Both methodologies are popular protocols, and while both protect against malicious nodes, PoS could be interpreted as more prone to attacks due to requiring lower miner costs and efforts [6]. On the other hand, PoW are considered to affect the scalability of the system in a negative manner [10]. Some lesser-known alternatives include Ripple, Practical Byzantine Fault Tolerance (PBFT), and Round Robin Consenses Protocols [13, 14].

Considering a case where MASs are used in sensitive domains, such as eHealth or Finance, data integrity is of increased importance. In addition to membership services, i.e., establishing the agent identity and certification of authenticity, management of identity is offered through BCT by means of Network Peers, as depicted in Figure 3, extracted from [15]. Moreover, Permissioned Blockchains, also known as *Private Blockchains* offer data privacy i.e., transactions are kept private to those who have permission for access. Permission for participation in the consensus is defined by the membership services where the user/node identities are also known [15], and in the case of MASs the agent identities would be known and verified. Whereas Permissionless Blockchains, *public blockchains*, offer the benefit of being fully decentralized, due to the anonymous and unlimited miners, this poses a threat to breach of privacy. Hence, the use of permissioned BCT is found to be more suitable for MAS application as the agents' identification assurance and management are a necessity, additionally ensuring reputation and authenticity to the system.

## IV. DISCUSSION ON INTEGRATION OF BCT TO MAS

The summation of how each BCT feature resolves each of the MAS vulnerabilities, defined in the previous section, portrayed in Figure 2 below, coloured in green. Depicting that the decentralized characteristic of BCT eliminates issues of leadership as well as the need for defining a trusted central authority, authority of agents and achieving consensus between agents. The BCT distributed ledger which provides a complete history that is immutable combats trust, data integrity and authenticity. These vulnerabilities are also solved through membership services of permissioned BCT, with the addition of improving privacy, agent synchronisation, formation, fault detection/prevention, agent verification and reputation. Whilst smart contracts avoid questioning trust and authenticity as conditions are defined with unanimous agreement and all executions are automatic. Agent leadership, formation, synchronisation would also be preassigned, reaching a consensus. Finally, consensus protocols are beneficial in areas of authenticity, authority, agent verification, establishing trust, fault prevention, leadership and reaching MAS consensus.

| MAS Vulnerabilities | BCT Features | | | | |
|---|---|---|---|---|---|
| | Decentralised System | Immutable History | Smart Contract | Consensus Protocols | Membership Services |
| Agent Verification | ✓ | ✓ | ✓ | ✓ | ✓ |
| Leadership | ✓ | ✓ | ✓ | ✓ | |
| Fault Detection | ✓ | ✓ | ✓ | ✓ | |
| Consensus | ✓ | ✓ | ✓ | ✓ | |
| Reputation | ✓ | ✓ | ✓ | ✓ | |
| Establishing Trust | ✓ | ✓ | ✓ | ✓ | |
| Privacy | ✓ | ✓ | | | ✓ |
| Authority | ✓ | ✓ | ✓ | ✓ | ✓ |
| Authenticity | ✓ | ✓ | ✓ | ✓ | ✓ |
| Data Integrity | | ✓ | | | |

**Figure 2 – Table Portraying which BCT Features are a solution to which MAS Vulnerabilities, represented by cells in green**

The consideration of permissioned vs. permissionless BCT application, can be dependent per application of the MAS, for example if permissioned BCT can be applied in scenarios where privacy is a more valued feature than transparency, such as in the eHealth, private financial transaction domain or household energy domain. And vice-versa for permissionless BCT in domains such as public health, publicly accessible governmental finances, and energy grids. Consensus protocols then vary according to this attribute of the Blockchain where protocols like PoW are for permissionless BCT applications and Round Robin Protocol (RRP) is applicable to permissioned BCT. The possibility of a *consortium* (hybrid) system could be considered in an effort to gain the privacy and control features of permissioned BCT while keeping the fully decentralised benefit of permissionless BCT.

Under technical consideration, the framework of MAS has had feasibly proposed architectures of integration with BCT. Where JADE (Java Agent Development Framework) Is the most commonly known and used platform for MASs, and for BCTs, the Hyperledger Fabric Network is the framework platform. Below portrays a chosen conceptual architecture for merging the frameworks of BCT and MAS, extracted from [15]. Where CA-A1 refers to a Certification Authority Agent, inhibiting an agent mind and ability to interact with the BCT Network, and BC-A represents regular MAS agents and associated abilities in addition to requiring certification (eCert) to the CA-Agent and execution of services including appending the ledger [15]. While remaining BCT dynamics include a Membership Service (ms.example.org), an ordering service (orderer.example.org) and 3 peers (peern.org1.example.org) It may also be noted that the framework stores IDs of agents and what services they perform.

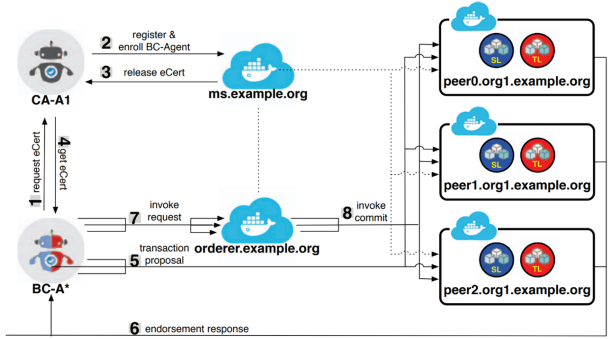

**Figure 3 - Example System Architecture of a BCT Integrated MAS System, extracted from [15]**

From an ethical perspective, concerns are additionally present, specifically regarding smart contracts. Where it has been suggested there is possibility of blockchain *fraud*, or *malicious smart contract developers* that might involve malicious behaviours of MAS agents' other parties may have not realised due to the complexity of certain systems [4, 5]. Additional technical challenges of merging BCT with MAS can be summarised as scalability, displaying real-time information, full disclosure of information, and traceability of the system.

## V. CONCLUSION AND FUTURE WORK

In summation, BCT offers multiple benefits for adoption into MAS, solving problems of key vulnerabilities, such as of security through its hash encryption in combination with multiple options of consensus protocols along with its distributed ledger and decentralized programme to name a few. This paper also explains the suitability of providing a solution to other MAS vulnerabilities such as identity management, through the use of permissioned BCT and Peer-to-Peet Networking as well as transaction guarantees through the use of smart contracts on a blockchain.

Continuing from the literature review conducted, future work of this paper entails producing a proposal of an improved integrated BCT with MAS System, including a comprehensive study implementing the chosen conceptual design () in a comparative manner, e.g., permissioned vs permissionless vs consortium BCT. As well as contrasting applicable consensus protocols weighing the advantages and disadvantages to provide a single 'best solution', and a real-world testing of application utilizing the system and its benefits compared to the current state of art.